\documentclass[11pt,a4paper]{article}

\usepackage{fullpage}
\usepackage{amsmath,amssymb}
\usepackage{hyperref}
\usepackage{graphicx}

\begin{document}

\renewcommand{\topfraction}{1} 
\renewcommand{\bottomfraction}{1}
\renewcommand{\floatpagefraction}{1}
\renewcommand{\textfraction}{0}


\newcommand{\ie}{i.e.,\ }
\newcommand{\eg}{e.g.,\ }

\newcommand{\const}{\operatorname{const.}} 


\newcommand{\rmd}{\,\mathrm{d}}

\newcommand{\Tr}{\operatorname{tr}}

\newcommand{\e}[1]{\operatorname{e}^{#1}}

\newcommand{\op}{\mathcal{O}}

\newcommand{\vev}[1]{\left\langle #1 \right\rangle}
\newcommand{\fvev}[1]{\langle #1 \rangle}
\newcommand{\jvev}[1]{\left\langle j\left| #1 \right| j \right\rangle}

\newcommand{\comb}[2]{\begin{pmatrix} #1\\#2\end{pmatrix}}

\newcommand{\Lag}{\mathcal{L}}
\newcommand{\Ham}{\mathcal{H}}

\newcommand{\Order}{\mathcal{O}}

\newcommand{\Mpl}{M_{\text{P}}}
\newcommand{\Lpl}{L_{\text{P}}}

\newcommand{\pois}[2]{\left\{#1,#2\right\}}
\newcommand{\dirac}[2]{\pois{#1}{#2}_D}
\newcommand{\commut}[2]{\left[#1,#2\right]}

\newcommand{\bx}{\mathbf{x}}
\newcommand{\by}{\mathbf{y}}
\newcommand{\bk}{\mathbf{k}}
\newcommand{\bv}{\mathbf{v}}

\newcommand{\Nq}{N_{(q)}}
\newcommand{\Nj}{N_{(j)}}
\newcommand{\Nper}{N_{(\perp)}}
\newcommand{\Np}{N_{(\parallel)}}

\newcommand{\htt}{\hat{h}}
\newcommand{\ptt}{\hat{\pi}}

\begin{center}

{\Large \textbf{Counting Photons in Static Electric and Magnetic Fields}}\\[2em]

\renewcommand{\thefootnote}{\fnsymbol{footnote}}
Wolfgang M{\"u}ck${}^{a,b}$\footnote[1]{wolfgang.mueck@na.infn.it}\\[2em]
\renewcommand{\thefootnote}{\arabic{footnote}}
${}^a$\emph{Dipartimento di Fisica, Universit\`a degli Studi di Napoli "Federico II"\\ Via Cintia, 80126 Napoli, Italy}\\[1em] 
${}^b$\emph{Istituto Nazionale di Fisica Nucleare, Sezione di Napoli\\ Via Cintia, 80126 Napoli, Italy}\\[2em]

\abstract{We describe the electromagnetic field by the massless limit of a massive vector field in the presence of a Coulomb gauge fixing term. The gauge fixing term ensures that, in the massless limit, the longitudinal mode is removed from the spectrum and only the two transverse modes survive. The system, coupled to a classical conserved current, is quantized in the canonical formalism. The classical field configurations due to time-independent electric charges and currents are represented by coherent states of longitudinal and transverse photons, respectively. The occupation number in these states is finite. In particular, the number of longitudinal photons bound by an electric charge $q$ is given by $N=q^2/(16\pi\hbar)$.
}

\end{center}

\section{Introduction}

Coherent states are known to provide the bridge between the classical and quantum dynamics of quantum mechanical systems and are important in a wide range of physical and mathematical applications \cite{Klauder}. Most physics students learn about them from text books \cite{Schiff,ItzyksonZuber}. In electrodynamics, they were introduced by the seminal work of Glauber \cite{Glauber:1963tx} and provide the basis for the theoretical understanding of infrared problems \cite{Chung:1965zza,Kulish:1970ut,Papanicolaou:1976sv}. The common wisdom is that the electromagnetic field generated by 
an external current, such as a moving point charge, contains an infinite number of \emph{physical}, but very soft, photons. Yet, to the best of my knowledge, the simple question of how many \emph{time-like} or \emph{longitudinal} photons are bound in a static electric field has neither been asked nor answered. Probably, this question was irrelevant for physical applications, such as the calculation of S-matrix elements, but recent developments have changed this status.

Classicalization \cite{Dvali:2010jz,Dvali:2010bf,Dvali:2011th} has been proposed as a mechanism, by which theories with certain non-renormalizable interactions can achieve ultra-violet completeness and prevent short distances from being probed by scattering experiments. Theories with such interactions are quite special in quantum field theory. Typically, they allow for backgrounds with superluminal propagation, because of which they cannot be ultra-violet completed in the usual Wilsonian sense \cite{Dvali:2012zc,Vikman:2012bx}. Briefly, the ultra-violet completeness is achieved through the formation of classical objects knows as \emph{classicalons}, which are similar to solitons. This mechanism is most efficient for gravity, where the typical classicalons are represented by black holes. Subsequently, it was proposed that a black hole of mass $M$ is nothing but a system of\footnote{This formula should be intended as valid up to a numerical factor of order unity.}
\begin{equation}
\label{intro:N}
	N = \frac{M^2}{M_P^2}
\end{equation}
gravitons at the verge of a quantum phase transition to a Bose Einstein condensate (BEC) \cite{Dvali:2011aa}. $M_P$ is the Planck mass.
This idea has been further explored in \cite{Dvali:2012rt,Dvali:2012gb,Dvali:2012en,Dvali:2012uq,Dvali:2012wq,Dvali:2013vxa,Flassig:2012re,Casadio:2013hja}. 
The statement of \eqref{intro:N} is that \emph{every} mass $M$ is surrounded by a gravitational field containing, on average, $N$ gravitons. If the size of the mass distribution becomes of about the size of its gravitational radius, \ie under the same conditions under which, classically, a black hole forms, then these $N$ gravitons undergo a quantum phase transition to a BEC. 

Because formulae similar to \eqref{intro:N} are central to many arguments in favour of the classicalization phenomenon, one must critically question the origin of this formula. In a previous paper \cite{Mueck:2013mha}, it was argued that the number $N$ can be calculated by considering a classical gravitational field as a coherent state of gravitons. The argument was based on the simpler case of an electric field around a charge $q$. Such a field was shown to contain about $q^2/(4\pi\hbar)$ photons (up to a numerical factor), and the analogy between Newtonian gravity and electrostatics was used to support \eqref{intro:N}. Recently, the problem of representing the static electric field of a point charge as a coherent state of unphysical photons was also studied in \cite{Barnich:2010bu}, with similar motivations. 

In the present paper, the question of how to count the photons in static electric and magnetic fields is considered in more detail, filling in the technical details of a consistent quantization from first principles that was missing in \cite{Mueck:2013mha}. In \cite{Mueck:2013mha}, it was argued that a small photon mass, intended as an infrared regulator to be removed at the end of the calculation, is needed for purely dimensional reasons in relation with the definition of a photon number operator from the electric potential. Introducing a photon mass gives rise to a third polarization for propagating fields, the longitudinal photon. Because of its feeble interaction with matter, it may not have observable consequences,\footnote{Placing limits on photon and graviton mass is an active field of experimental research. For recent reviews on this subject, see \cite{Tu:2005,Goldhaber:2008xy}.} but its analogue for massive (Fierz-Pauli) tensor fields gives rise to the van~Dam-Veltman-Zakharov discontinuity in the massless limit \cite{vanDam:1970vg,Zakharov:1970cc}. Hence, if possible, we should construct a theory of photons with rest mass such that, in the massless limit, only the two transverse degrees of freedom survive. This requirement essentially rules out theories derived from the Proca Lagrangian, in which all three polarizations appear on an equal footing. 

In what follows, we argue that canonical quantization of a massive vector field with a gauge fixing term corresponding to the Coulomb gauge, $\partial_i A^i(\bx)=0$, is a suitable approach. It has the disadvantage that it is not relativistically invariant, but one should bear in mind that explicitly specifying a non-zero source breaks this invariance anyway. 
There are several properties in its favour. First, even classically, the Coulomb gauge is more restrictive than the covariant Lorentz gauge, which is reflected in the fact that it does not remove the primary constraint. Second, adding a mass term as an infrared regulator makes the constraints second-class, implying that quantization can be done in a standard fashion with a positive-definite-norm Hilbert space. There are three propagating degrees of freedom, two transverse and a longitudinal mode, with different dispersion relations. In the massless limit, the energy of the longitudinal mode is pushed to infinity, so that this mode does not contribute to the partition function. Hence, the massless limit truly yields the dynamics of the massless vector field. The ground state in the presence of a static classical source is shown to be a coherent state with finite occupation number.

The remainder of the paper is organized as follows. In Sec.~\ref{em:lagrangian}, the Lagrangian is introduced and it is shown that the Green's function is such that, in the massless limit, only two modes with the dispersion relation of a massless vector field survive. The canonical quantization of the system is carried out in Sec.~\ref{em:quant}. In Sec.~\ref{em:vac}, the ground state in a presence of static sources is identified as a coherent state, and the occupation number is calculated. 
Sec.~\ref{conc} contains the conclusions.

\section{Lagrangian and classical field equations}  
\label{em:lagrangian}
Following the discussion in the introduction, we consider the Lagrangian density\footnote{The metric signature is $(-+++)$, Greek indices run from 0 to 3, Latin ones from 1 to 3. The field strength tensor is $F_{\mu\nu}= \partial_\mu A_\nu -\partial_\nu A_\mu$, and $m$ denotes mass$/\hbar$ with units of inverse length.}
\begin{equation}
\label{em:lag}
	\Lag = -\frac14 F_{\mu\nu}F^{\mu\nu} -\frac12 m^2 A_\mu A^\mu -\frac12 \sigma \left(\partial_i A^i \right)^2 
	-j^\mu A_\mu~.
\end{equation}
The dimensionless parameter $\sigma$ ensures that the classical fields satisfy the Coulomb gauge condition in the massless limit. The external, classical current, $j^\mu$, in addition to being conserved, will be assumed to be time-independent,
\begin{equation}
\label{em:j.stationary}
	\partial_\mu j^\mu=0~,\qquad \partial_0 j^\mu=0~.
\end{equation}
Therefore, it describes a static charge distribution $j^0(\bx)$ and a stationary, divergence-free current field $j^i(\bx)$, $\partial_i j^i(\bx)=0$. For the sake of generality, we shall ignore \eqref{em:j.stationary} for the time being and use it at the appropriate moment. 

The classical field equations are
\begin{equation}
\label{em:eoms}
	\left[ \left(\square -m^2\right) \delta^\mu_\nu - \partial^\mu \partial_\nu
	+ \sigma \delta^\mu_i \delta_\nu^j \partial^i \partial_j \right] A^\nu = j^\mu~.
\end{equation}

The connected two-point function for the vector fields is obtained by inverting the differential operator in \eqref{em:eoms},
\begin{equation}
\label{em:2point}
	\vev{A_\mu(x)A_\nu(0)} = \frac1{\square-m^2}\left[ \eta_{\mu\nu} 
	- \frac{(\square-m^2 +\sigma\triangle)\partial_\mu\partial_\nu
	 -2 \sigma\triangle \delta_{(\mu}^i \partial^{}_{\nu)} \partial_i 
	 + \sigma m^2 \delta_\mu^i \delta_\nu^j \partial_i \partial_j}{m^2(\square-m^2+ \sigma\triangle) -\sigma\triangle^2} 
	 \right] \delta^4(x)~,
\end{equation}
where $\triangle = \partial_i \partial^i$ and $\square = \partial_\mu \partial^\mu$, and parentheses around indices indicate their symmetrization. Obviously, as long as $\sigma\neq 0$, the limit $m\to 0$ is non-singular. In particular, when coupled to a conserved current, the limit of the second term in the brackets reduces to a pure gauge term, which does not contribute to the field strength. The trace of the residue of the pole at $\square=m^2$ is 2, implying that this pole arises from two degrees of freedom, which we shall identify below as the transverse modes.  
Morever, it is instructive to translate the denominator of the second term into 4-d momentum space,
\begin{equation}
\label{em:denominator}
	m^2 \left(\square-m^2+ \sigma\triangle \right) -\sigma\triangle^2 \Rightarrow
	m^2 \left[ (k_0)^2 -\frac{\omega_k^2 \tilde{\omega}_k^2}{m^2}\right]~,
\end{equation}
with 
\begin{equation}
\label{em:omega}
	\omega_k = \sqrt{m^2+k^2}~,\qquad \tilde{\omega}_k = \sqrt{m^2+\sigma k^2}~, \qquad k^2 = k_i k^i~.
\end{equation}
This shows that, for $\sigma> 0$, the energy of the third mode is pushed to infinity in the massless limit.

\section{Canonical quantization}
\label{em:quant}
The Lagrangian \eqref{em:lag} can be straightforwardly quantized in the canonical formalism, if one takes care to treat the second-class constraints using Dirac's prescription \cite{Henneaux}. Let us go through the quantiziation procedure in detail.
The canonical momenta, which formally satisfy the Poisson brackets 
\begin{equation}
\label{em:poisson}
	\pois{A_\mu(\bx)}{\pi^\nu(\by)} = \delta_\mu^\nu\, \delta^3(\bx-\by)~,
\end{equation}
are 
\begin{equation}
\label{em:momenta}
	\pi^0 = 0~,\qquad \pi^i =F^{i0}~.
\end{equation}
The Hamiltonian density is
\begin{equation}
\label{em:ham}
	\Ham = \frac12 \pi^i\pi^i + \frac14 F_{ij}F^{ij} +\frac12 m^2 A_i A_i 
	+\frac12 \sigma \left(\partial_i A^i \right)^2 +j^i A_i -\frac12 m^2 (A_0)^2+\left(j^0-\partial_i \pi^i\right)A_0~.
\end{equation}
The first equation of \eqref{em:momenta} is a primary constraint. Consistency leads to a secondary constraint, and both constraints are second-class for non-zero $m$,
\begin{align}
\label{em:constraint1}
	\phi_1 &= \pi^0 \approx 0~,\\
\label{em:constraint2}
	\phi_2 &= m^2 A_0 -j^0 +\partial_i \pi^i \approx 0~,\\
\label{em:pois.cons}
	\pois{\phi_1(\bx)}{\phi_2(\by)} &= -m^2\,  \delta^3(\bx-\by)~.
\end{align}
After constructing the Dirac bracket
\begin{equation}
\label{em:dirac}
	\dirac{F}{G} = \int\rmd^3x\,\frac{\delta F}{\delta A_i(\bx)}
	\left[\frac{\delta G}{\delta \pi^i(\bx)} 
	+ \frac1{m^2} \frac{\partial}{\partial x^i} \frac{\delta G}{\delta A_0(\bx)} \right]
	- ( F \leftrightarrow G)~,
\end{equation}
the constraints \eqref{em:constraint1} and \eqref{em:constraint2} can be imposed strongly. One could use the constraint \eqref{em:constraint2} to eliminate $A_0$, so that the Dirac bracket \eqref{em:dirac} of the variables $(A_i, \pi^i)$ would coincide with the Poisson bracket. Instead, we shall eliminate the longitudinal modes of $\pi^i$ and continue to use $A_0$ as a field variable. The reason for this choice is that we want the field vacuum to satisfy $\fvev{0|A_\mu|0}=0$. Details will be discussed further below.

To proceed, we transform to (3-d) momentum space\footnote{Formally, replace $\partial_i\to -i k_i$. The integration measure in momentum space is $\rmd^3k/(2\pi)^3$, where $k$ has units of inverse length. Notice that this is not the relativistically invariant measure, to which the reader may switch with the appropriate changes.} and separate the vectors into their transverse and longitudinal components,
\begin{align}
\label{em:A.long.trans}
	A_i(\bk) &= A_{\perp i}(\bk) + i \frac{k_i}{k} A_{\parallel}(\bk)~, & A_{\parallel}(\bk) &= -i \frac{k^i}{k} A_i(\bk)~, \\
\label{em:pi.long.trans}	
	\pi^i(\bk) &= \pi^i_{\perp}(\bk) + i \frac{k^i}{k} \pi_{\parallel}(\bk) ~, & \pi_{\parallel}(\bk) &= -i \frac{k_i}{k} \pi^i(\bk)~, 
\end{align}
where $k= \sqrt{k_i k^i}$. Reality of the vector fields in position space implies
\begin{equation}
\label{em:reality}
	A_{\perp i}(-\bk) = A_{\perp i}(\bk)^\ast~,\quad  A_{\parallel}(-\bk) = A_{\parallel}(\bk)^\ast~, \quad 
	\pi^i_{\perp}(-\bk) = \pi^i_{\perp}(\bk)^\ast~,\quad  \pi_{\parallel}(-\bk) = \pi_{\parallel}(\bk)^\ast~,
\end{equation}
and similarly for the external current.

Hence, after eliminating $\pi_{\parallel}$ by means of \eqref{em:constraint2}, the Hamiltonian density \eqref{em:ham} becomes, in momentum space, 
\begin{align}
\label{em:ham1}
	\Ham &= \Ham_\perp + \Ham_\parallel~,\\
\label{em:ham2}
	\Ham_\perp &= \frac12 |\pi_\perp|^2 +\frac12 \omega_k^2 \left|A_\perp +\frac1{\omega_k^2} j_\perp\right|^2
		-\frac1{2\omega_k^2}|j_\perp|^2~,\\
\label{em:ham3}
	\Ham_\parallel &= \frac{m^2\omega_k^2}{2k^2} \left|A_0 -\frac1{\omega_k^2} j^0\right|^2 
	+\frac12 \tilde{\omega}_k^2 \left|A_\parallel +\frac1{\tilde{\omega}_k^2} j_\parallel\right|^2 
	+\frac1{2\omega_k^2}|j^0|^2 -\frac1{2\tilde{\omega}_k^2} |j_\parallel|^2~,
\end{align}
where $\omega_k$ and $\tilde{\omega}_k$ were defined in \eqref{em:omega} and the current has been decomposed in an obvious way. The non-zero Dirac brackets \eqref{em:dirac} of the remaining fields are
\begin{align}
\label{em:dirac2}
	\dirac{A_{\perp i}(\bk)}{\pi_\perp^j(\bk')} &= P_i^j\, (2\pi)^3\, \delta^3(\bk+\bk')~,\\
\label{em:dirac3}
	\dirac{A_0(\bk)}{A_\parallel(\bk')} &= \frac{k}{m^2} (2\pi)^3\, \delta^3(\bk+\bk')~,
\end{align}
where $P_i^j=\delta_i^j -k_ik^j/k^2$ is the transverse projector. 

After quantization, the fields can be expressed in terms of the usual ladder operators\footnote{The ladder operators satisfy $\commut{a(\bk)}{a^\dagger(\bk')}=(2\pi)^3 \delta^3(\bk-\bk')$. For the transverse modes, the commutator involves also the transverse projector.}
\begin{align}
\label{em:Api.quant}
	A_{\perp i}(\bk) &= i \sqrt{\frac{\hbar}{2\omega_k}} \left[ a_{\perp i}(\bk) - a^\dagger_{\perp i}(-\bk) \right]~,  &
	\pi_{\perp i}(\bk) &= \sqrt{\frac{\hbar\omega_k}{2}} \left[ a_{\perp i}(\bk) + a^\dagger_{\perp i}(-\bk) \right]~,\\
\label{em:A0.quant}
	A_0(\bk) &= \sqrt{\frac{\hbar k^2 \tilde{\omega}_k}{2m^3\omega_k}}  \left[ a(\bk) + a^\dagger(-\bk) \right]~, &
	A_\parallel(\bk) &= -i \sqrt{\frac{\hbar \omega_k}{2m\tilde{\omega}_k}}  \left[ a(\bk) - a^\dagger(-\bk) \right]~. 
\end{align}	
Finally, the (normal-ordered) Hamiltonian density \eqref{em:ham1}--\eqref{em:ham3} becomes
\begin{equation}
\label{em:ham4}
	\Ham = \hbar \omega_k \left[ a'_{\perp i}{}^\dagger(\bk) a'_{\perp i}(\bk) 
		+ \frac{\tilde{\omega}_k}{m} a'{}^\dagger(\bk) a'(\bk) \right] +\Ham_j~,
\end{equation}
where the primed ladder operators are related to the unprimed ones by a bosonic shift,
\begin{align}
\label{em:shift1}
	a'_\perp(\bk) &= a_\perp(\bk) -i \sqrt{\frac1{2\hbar \omega_k^3}}\, j_\perp(\bk)~,\\
\label{em:shift2}
	a'(\bk) &= a(\bk) -\sqrt{\frac{m^3}{2\hbar k^2 \omega_k^3 \tilde{\omega}_k}}\, j^0(\bk)
		+i \sqrt{\frac{m}{2\hbar \omega_k \tilde{\omega}_k^3}}\, j_\parallel(\bk)~,
\end{align}
and $\Ham_j$ denotes the ground state energy density,
\begin{equation}
\label{em:ham.j}
	\Ham_j = \frac1{2\omega_k^2} \left( |j^0(\bk)|^2 -|j_\perp(\bk)|^2 
	-\frac{\omega_k^2}{\tilde{\omega}_k^2} |j_\parallel(\bk)|^2 \right)~.
\end{equation}

This completes the quantization procedure. Before investigating the properties of the ground state, let us briefly discuss the importance of the parameter $\sigma$, which is hidden in $\tilde{\omega}_k=\sqrt{m^2+\sigma k^2}$. As is evident from \eqref{em:ham4}, without the gauge fixing term ($\sigma=0$), all three oscillators would have the same energy, in agreement with the massive vector field of the Proca Lagrangian. Instead, in the presence of the Coulomb gauge fixing term, \ie for positive $\sigma$ ($\sigma<0$ would lead to imaginary energy eigenvalues), the longitudinal mode is physically distinct from the two transverse modes, and the limit $m\to 0$ removes it from the spectrum by pushing its energy eigenvalue to infinity. This agrees with our findings in Sec.~\ref{em:lagrangian}. Therefore, explaining the absence of physically observable consequences of this kind of longitudinal photon does not need to invoke its feeble coupling to matter. 
The choice $\sigma=1$ is particularly appealing, because $\tilde{\omega}_k =\omega_k$ leads to the relativistically invariant combination $-j_\mu(-\bk) j^\mu(\bk)$ in the ground state energy density \eqref{em:ham.j}. In the next section, we will give another argument in favour of this choice.

\section{Ground state as a coherent state}
\label{em:vac}

Formally, the source-dependent ground state $|j\rangle$ of the Hamiltonian \eqref{em:ham4} is given by 
\begin{equation}
\label{em:vacuum}
	a'(\bk) |j\rangle = 0~,
\end{equation}
where the annihilation operators of all three components are intended for brevity. 
The field vacuum $|0\rangle$, however, is defined by $a(\bk)|0\rangle=0$ and obviously satisfies $\vev{0|A_\mu(\bk)|0}=0$. In terms of $|0\rangle$, $|j\rangle$ is a coherent state,\footnote{We work with normalized states, $\vev{j|j}=1$.} 
\begin{equation}
\label{em:coherent}
	a(\bk) |j\rangle = \alpha(\bk) |j\rangle~,\qquad 
	|j\rangle = \e{-\frac12 N} \e{\int \frac{\rmd^3 k}{(2\pi)^3} \alpha(\bk) a^\dagger(\bk)} |0\rangle~,
\end{equation}
where the coefficients $\alpha(\bk)$ can be read off from \eqref{em:shift1}--\eqref{em:shift2}, and $N$ denotes the expectation value of the number of field quanta (photons) in the state $|j\rangle$,
\begin{align}
\notag 
	N&= \int\frac{\rmd^3 k}{(2\pi)^3} \vev{j\left|a^\dagger(\bk) a(\bk)\right|j} = \int\frac{\rmd^3 k}{(2\pi)^3} |\alpha(\bk)|^2 \\
\label{em:N}
	&=  \int\frac{\rmd^3 k}{(2\pi)^3} \frac1{2\hbar\omega_k^3} \left[ \frac{m^3}{k^2\tilde{\omega}_k} |j^0|^2 
	  +\frac{m\omega_k^2}{\tilde{\omega}_k^3} |j_\parallel|^2 + |j_\perp|^2 \right]~.
\end{align}

This formal procedure has a caveat. The interpretation of $N$ as the number of coherent quanta is only valid, if it is finite, because only then is the bosonic shift \eqref{em:shift1}--\eqref{em:shift2} unitary. Indeed, $\vev{0|j}=\e{-N/2}$ implies that the Fock spaces built with the primed and unprimed oscillators, respectively, are orthogonal to each other, if $N$ diverges. A very nice discussion of this can be found in Chapter~4 of \cite{ItzyksonZuber}. 
However, as we will show shortly, $N$ is a finite number in the case of time-independent sources, even in the massless limit.

Consider a static distribution of point charges, 
\begin{equation}
\label{em:qx.system}
	j^0(\bx) = \sum\limits_n q_n \delta(\bx-\bx_n)~,\qquad \sum\limits_n q_n =q~,
\end{equation}
and a stationary current field $j^i(\bx)$, which must be of zero divergence because of \eqref{em:j.stationary}, $\partial_i j^i(\bx)=0$. 
The number of coherent photons \eqref{em:N} splits into two contributions, $N=\Nq+\Nj$, corresponding to the first and the last terms in the bracket in the integrand, respectively.
For $\Nq$, one has
\begin{align}
\label{em:Nq}
	\Nq &= \sum\limits_{n,l} \frac{q_nq_l}{2\hbar(2\pi)^3} \int\rmd^3 k\frac{m^3}{k^2\tilde{\omega}_k \omega_k^3} 
	\e{i\bk\cdot(\bx_n-\bx_l)} \\ 
\label{em:Nq1}
	&= \sum\limits_{n,l} \frac{q_nq_l}{2\hbar(2\pi)^3} \int\rmd^3 \kappa\, \kappa^{-2} (1+\kappa^2)^{-3/2}(1+\sigma\kappa^2)^{-1/2}
		\e{im\vec{\kappa}\cdot(\bx_n-\bx_l)}~,
\end{align}
where the integration variables have been rescaled by the mass parameter to obtain \eqref{em:Nq1}. Thus, $m$ appears only in the exponent and, because the integrand of the radial integral goes as $\kappa^{-4}$ for large $\kappa$, the limit $m\to0$ can be taken before integrating. After doing so, the charges can be summed up, and the integration over the angular variables yields
\begin{align}
\notag
	\Nq &= \frac{q^2}{2\hbar(2\pi)^2} \int\limits_0^\infty \rmd z\, z^{-1/2} (1+z)^{-3/2}(1+\sigma z)^{-1/2} \\
\label{em:Nq2}
	 &= \frac{q^2}{16\pi \hbar} \mathrm{F}\left(\frac12,\frac12;2;1-\sigma\right)~,
\end{align}
where $\mathrm{F}$ denotes a Gauss hypergeometric function. With some hindsight, one may do the trick of setting $\bk=m\vec{\kappa}$ directly in the momentum space integral of \eqref{em:N}, for the term involving $j^0$, so that $m$ appears only in $|j^0(m\vec{\kappa})|^2$. As the integral is convergent in the UV, the limit $m\to 0$ can be done before integrating and gives rise to $|j^0(0)|^2=q^2$, which is pulled out of the momentum integral. 

For $\sigma=1$, the result \eqref{em:Nq2} is simply
\begin{equation}
\label{em:Nq3}
	\Nq = \frac{q^2}{16\pi \hbar}~.
\end{equation}
One can make the interesting observation that the result \eqref{em:Nq3} does not change, if one replaces the fraction $\frac{m^2}{k^2}$ in the integrand of \eqref{em:Nq} by unity. Following the same steps as above, one obtains
\begin{align}
\label{em:Nq.mod}
	\Nq' &= \sum\limits_{n,l} \frac{q_n q_l}{2\hbar(2\pi)^3} \int\rmd^3 k\frac{m}{\tilde{\omega}_k \omega_k^3} 
	\e{i\bk\cdot(\bx_n-\bx_l)} \\ 
\label{em:Nq1.mod}
	&= \frac{q^2}{16\pi\hbar} \mathrm{F}\left(\frac12,\frac32;2;1-\sigma\right)~.
\end{align}
Obviously, for $\sigma=1$, this yields $\Nq'=\Nq$, while the hypergeometric function in \eqref{em:Nq1.mod} diverges for $\sigma=0$. In addition, for $\sigma=1$, the modified formula \eqref{em:Nq.mod} can be rewritten as the local expression
\begin{equation}
\label{em:N.mod}
	\Nq= \Nq' = \frac{m}{2\hbar} \int \rmd^3 x |A_{0,cl}(\bx)|^2~,
\end{equation}
where $A_{0,cl}$ is the classical field solution given by $\omega_k^2 A_{0,cl}(\bk)= j^0(\bk)$. This is another reason to set $\sigma=1$. The result \eqref{em:N.mod} confirms the heuristic approach taken in \cite{Mueck:2013mha} and also fixes the numerical factor that was left undetermined. 

From \eqref{em:N}, the occupation number associated with the current field, $j^i(\bx)$, is 
\begin{equation}
\label{em:Nj}
	\Nj = \frac1{2\hbar} \int \rmd^3x \rmd^3y\, \vec{\jmath}(\bx)\cdot \vec{\jmath}(\by) \int \rmd^3k 
	\frac{\e{i\bk\cdot(\bx-\by)}}{(2\pi\omega_k)^3}~.
\end{equation}
Doing the $k$-integral leads to  
\begin{equation}
 \label{em:Nj2}
	\Nj = \frac1{(2\pi)^2\hbar} \int \rmd^3x \rmd^3y\, \vec{\jmath}(\bx)\cdot \vec{\jmath}(\by)\, \mathrm{K}_0(m|\bx-\by|)~,
\end{equation}
where $\mathrm{K}_0$ is a modified Bessel function. Letting $m\to0$ in this expression would seem to lead to a logarithmic singularity, but we should remember that the current has zero divergence. Indeed, substituting  
\begin{equation}
\label{em:trick}
	\delta_{ij} = \delta_{ij} - \frac{(x-y)_i(x-y)_j}{|\bx-\by|^2} 
	+ \frac{(x-y)_i}{|\bx-\by|} \frac{\partial}{\partial x^j} |\bx-\by|
\end{equation}
into \eqref{em:Nj2} and integrating the last term by parts yields
\begin{equation}
 \label{em:Nj3}
	\Nj = \frac{m}{(2\pi)^2\hbar} \int \rmd^3x \rmd^3y\, j^i(\bx) j^j(\by) \frac{(x-y)_i(x-y)_j}{|\bx-\by|}
	\,\mathrm{K}_1(m|\bx-\by|)~.
\end{equation}
Now, the limit $m\to0$ can be taken, and the final result is
\begin{equation}
\label{em:Nj4}
	\Nj = \frac1{(2\pi)^2\hbar} \int \rmd^3x \rmd^3y\, j^i(\bx) j^j(\by) \frac{(x-y)_i(x-y)_j}{|\bx-\by|^2}~.
\end{equation}
This is a finite, configuration-dependent quantity for non-singular current fields $j^i(\bx)$. For example, for a circular  current, $\vec{\jmath}(\bx) = J\,\delta(z)\,\delta(r-R)\, \mathbf{e}_\phi$, where $(z,r, \phi)$ denote $\bx$ in cylindrical coordinates, one obtains $\Nj =\frac{(JR)^2}{2 \hbar}$. 
Note that the photons bound by the current field have transverse polarization. As a consequence, their number does not depend on the gauge parameter $\sigma$.

To conclude this section, let us verify that the expectation values of all fields in the ground state $|j\rangle$ coincide with the corresponding classical fields. This is done by substituting the field shifts \eqref{em:shift1}--\eqref{em:shift2} into the expressions \eqref{em:Api.quant}--\eqref{em:A0.quant} and using \eqref{em:vacuum}. The results are
\begin{align}
\label{em:Api.t.cl}
	\jvev{A_{\perp i}(\bk)} &= -\frac1{\omega_k^2} j^i(\bk)= A_{i,cl}(\bk)~, & \jvev{\pi_{\perp i}} &=0~,\\
\label{em:A0Al.cl}
	\jvev{A_0(\bk)} &=\frac1{\omega_k^2} j^0(\bk) = A_{0,cl}(\bk)~, & \jvev{A_\parallel} &=0~,
\end{align}
where $A_{\mu,cl}$ is the static classical field solution, as one can easily verify. For the magnetic field strength, we simply have
\begin{align}
\notag
	\jvev{F_{ij}(\bk)} &= -i \jvev{k_i A_j(\bk) -k_j A_i(\bk)} = -i \left[k_i \jvev{A_j(\bk)} - k_j\jvev{A_i(\bk)} \right] \\
\label{em:Fmag.cl}
	&= -i \left[ k_i A_{j,cl}(\bk) -k_j A_{i,cl}(\bk) \right] = F_{ij,cl}(\bk)~.
\end{align}
The interesting quantity is the electric field strength, 
\begin{equation}
\label{em:F.cl}
	\jvev{F^{i0}(\bk)} = \jvev{\pi^i(\bk)} = \jvev{\pi_\perp^i(\bk)+i\frac{k^i}{k} \pi_\parallel}~.
\end{equation}
Eliminating $\pi_\parallel$ by the constraint \eqref{em:constraint2} and then using \eqref{em:Api.t.cl} and \eqref{em:A0Al.cl}, one obtains as expected
\begin{equation}
\label{em:F.cl2}
	\jvev{F^{i0}(\bk)} = i \frac{k^i}{k^2} \left[j^0(\bk)- m^2 A_{0,cl}(\bk) \right] = ik^i A_{0,cl}(\bk) 
	=F^{i0}_{cl}(\bk)~.
\end{equation}

\section{Conclusions}
\label{conc}

We have shown that the Lagrangian \eqref{em:lag}, describing the photon field as a vector field with a Coulomb gauge fixing term and a small mass to regularize the infrared behaviour of the theory, can be quantized in the standard canonical formalism. The massless limit of this theory is non-singular and contains only the transverse modes of the photon field. The propagating longitudinal modes freeze in this limit, their mass being pushed to infinity. Nevertheless, these modes should not be regarded as entirely unphysical, because they are responsible for the classical Coulomb field around electric charges. A quantum description of the Coulomb field in terms of a coherent state of longitudinal photons has been given, and it has been shown that the occupation number in this state is finite and proportional to the square of the total charge, supporting the  analogous value \eqref{intro:N} for gravitons. The photon number is, in principle, dependent on the parameter $\sigma$, but two aesthetic arguments favour the choice $\sigma=1$: First, the ground state energy \eqref{em:ham.j} takes a relativistically invariant form and, second, the occupation number \eqref{em:Nq3} can be expressed in the local form \eqref{em:N.mod}. Stationary electric currents give rise to a coherent state of transverse photons, with a finite occupation number given by \eqref{em:Nj4}. In the ground state, Ehrenfest's theorem is satisfied for all fields.

Some further comments on the dependence of $\Nq$ on $\sigma$ are due. At first sight, this dependence may appear puzzling, because $\sigma$ is the parameter of the Coulomb gauge fixing term, while physical quantities must not depend on the choice of gauge. Furthermore, the classical field solution $A_{0,cl}$ does not involve $\sigma$, so how does $\sigma$ enter the corresponding coherent state? The answer to the first point is that the massive theory does not possess a gauge invariance. Therefore, strictly speaking, $\sigma$ is not a gauge parameter. Instead, different choices of $\sigma$ correspond to physically distinct theories, which differ precisely in the details of the longitudinal quanta that freeze in the massless limit. The second point can be understood from \eqref{em:ham3}, \eqref{em:dirac3} and \eqref{em:A0.quant}. Although the classical field $A_{0,cl}$ does not know about $\sigma$, the quantum dynamics of the canonical pair of variables $(A_0, A_\parallel)$ depends on it.  

The drawback of the formalism is that the result is not relativistically invariant. In fact, considering a uniformly moving charge distribution and calculating the occupation number \eqref{em:N}, one finds the following. First, the contribution due to the electric charge $j^0$ remains unchanged. Second, there is a finite contribution proportional to the square of the velocity from the term involving $j_\parallel$, but it may be removed by adding a total time derivative to the Lagrangian. More precisely, one can use the current conservation \eqref{em:j.stationary} to eliminate $j_\parallel$ from \eqref{em:lagrangian} in favour of $\partial_0 j^0$. Integrating by parts with respect to time and dropping the boundary term gives rise to a shift in the canonical momentum $\pi_\parallel$, such that $j^0$ disappears from \eqref{em:constraint2}, but appears in the $\parallel$ component of \eqref{em:momenta}. The subsequent analysis yields exactly the same results, except that $j_\parallel$ is absent. Last, the number of transverse photons due to the term with $j_\perp$ is divergent in the massless limit. Physically, this divergence can be understood as part of the Bremsstrahlung associated with the acceleration of the charge from rest to the velocity that is considered. There is, however, a formal reason for this divergence, which was mentioned as the caveat after \eqref{em:N}. The explicit time dependence of the Hamiltonian introduced by time-dependent sources implies that there is no state, which is stable under time evolution. In particular, the formal ground state $|j_0\rangle$ defined at the time $t_0$ will evolve into a coherent state which is physically different, at a later time $t_1$, from the formal ground state $|j_1\rangle$ defined at $t_1$.\footnote{The property of coherence is conserved under time evolution \cite{Klauder}. Note that we work in the Schr\"odinger picture. Hence, \emph{physically different} means that the two states differ by more than just a time-dependent phase.} This implies that the Fock bases built upon $|j_0\rangle$ and $|j_1\rangle$ cannot be related to each other by a unitary transformation. Infinite $N$ agrees with this conclusion. 

It would be important to rederive the results obtained in this paper using a regularization of the infrared divergences of the massless theory, which does not involve a mass term. Maybe, this would shed light on the peculiar dependence on $\sigma$ of our result. For example, one may consider a system in a finite volume, in which case boundary terms may become relevant. Remember that the analysis in momentum space systematically drops boundary terms by relying on the standard assumptions on the asymptotic behaviour of the fields for large distances, which are satisfied in the massive case. Another implication of zero mass is the presence of first-class constraints. Such a treatment would be particularly interesting in the prospect of considering the case of gravity, where the Fierz-Pauli mass term is not sufficient to render all the constraints second-class.

\section*{Acknowledgements}
Sincere thanks are due to Glenn Barnich and Paolo Aniello for useful discussions. Thanks also to the Universit\'e Libre de Bruxelles for kind hospitality. This research was supported in part by INFN, research initiative TV12.

\bibliographystyle{JHEP}
\bibliography{classicalization,photon_num}
\end{document}